\documentclass[aps,pra,twocolumn,superscriptaddress,showpacs]{revtex4-1}

\usepackage[dvips]{graphics}
\usepackage{graphicx}
\usepackage{amsfonts}
\usepackage{amssymb}
\usepackage{amsmath}
\usepackage{subfigure}
\usepackage{color}

\def\dis{\displaystyle}
\def\({\left(}
\def\){\right)}
\def\ls{\left[}
\def\rs{\right]}
\def\ga{\gamma_a}
\def\gp{\gamma_p}
\def\gs{\gamma_s}
\def\Ep{\overline{E}_+}
\def\Em{\overline{E}_-}
\def\ep{{E}_+}
\def\em{{E}_-}
\def\a{\alpha}
\def\alp{\alpha}
\def\k{\kappa}
\def\w{\omega}
\def\g{\gamma}
\def\ba{\begin{align}}
\def\ea{\end{align}}
\def\be{\begin{equation}}
\def\ee{\end{equation}}
\def\M{\mathcal{M}}
\def\v{\underline{v}}
\def\l{\lambda}

\begin{document}

\title{The spin-flip model of spin-polarized 
vertical-cavity surface-emitting lasers: asymptotic analysis, numerics, and experiments}
\author{H.\ Susanto}
\affiliation{Department of Mathematical Sciences, University of Essex, Wivenhoe Park, Colchester, CO4 3SQ, United Kingdom}
\email{hsusanto@essex.ac.uk}
\author{K.\ Schires}
\affiliation{Universit\'e Paris-Saclay, T\'el\'ecom ParisTech, CNRS LTCI, 46 rue Barrault, 75013 Paris, France}
\author{M.J.\ Adams}
\author{I.D.\ Henning}
\affiliation{School of Computer Engineering and Electronics Engineering, University of Essex, Wivenhoe Park, Colchester, CO4 3SQ, United Kingdom}

\begin{abstract}
The spin-flip model describing optically pumped spin-polarized vertical-cavity surface-emitting lasers is considered. The steady-state solutions of the model for elliptically-polarised fields are studied. Asymptotic analysis for the existence and stability of the steady-state solutions is developed, particularly in the presence of pump polarisation ellipticity. The expansion is with respect to small parameters representing the ellipticity and the difference between the total pump power and the lasing threshold. The analytical results are then confirmed numerically, where it is obtained that generally one of the steady-state solutions is stable while the other is not. The theoretical results are shown to be in qualitative agreement with the experiments. 

\end{abstract}

\maketitle

\section{Introduction}\label{intr}

Spin-polarized vertical-cavity surface-emitting lasers (VCSELs) offer advantages over conventional lasers such as threshold reduction, independent control of output polarization and intensity, and faster dynamics \cite{r1}. These features are a consequence of a spin-polarized electron population which can be achieved either by electrical injection using magnetic contacts or by optical pumping using circularly polarized light. Since the development of the electrically pumped spin-VCSELs \cite{ra1,ra2} and the presentation of the first electrically pumped spin-laser at room temperature \cite{ra3} at the latest, it is clear that
spin-lasers representing a promising new class of applicable room temperature spintronic devices beyond magnetoresistive effects. New applications are foreseen in optical information processing and data storage, optical communication, quantum computing and bio-chemical sensing (including chiral spectroscopy).

Various forms of instability are predicted to occur in spin-VCSELs, including periodic oscillations, polarisation switching and chaotic dynamics \cite{r2}. Triggerable, ultrafast (11.6 GHz) circular polarization oscillations that decay in a few nanoseconds have been experimentally observed in an 850 nm VCSEL with hybrid excitation (D.C.\ electrical plus pulsed circularly-polarized optical pumping) \cite{r3}. Self-sustained periodic oscillations that can be tuned from 8.6 to 11 GHz with the pump polarization have been reported for an optically pumped 1300 nm dilute nitride spin-VCSEL \cite{r4}. Simulations using the spin-flip model (SFM) \cite{r16} yielded good agreement with the latter experimental results \cite{r2,r5,r6}, confirming that the oscillation frequency is dominated by the birefringence of the active material in combination with the dichroism and spin relaxation processes, as originally predicted by Gahl et al \cite{r7}.

A widely-used test for spin-VCSEL behaviour is to measure the variation of output polarization ellipticity when that of the optical pump is varied from left circularly polarized (LCP) to right circularly polarized (RCP). Polarization “gain” is found in some cases when the output ellipticity exceeds that of the pump \cite{r1,r8,r9}. However, numerical simulations also indicate situations where switching can occur between opposite polarization states, i.e. from LCP to RCP output or vice versa, in spin-VCSELs with either quantum well \cite{r2,r6,r9} or quantum dot \cite{r10} active regions. Experimental results on dilute nitride quantum well spin-VCSELs have confirmed the existence of this polarization switching \cite{r11}. In order to understand this phenomenon, particularly the polarization selection mechanism(s), it is necessary to determine the regions of stability and of switching by performing a stability analysis as a function of pump strength and polarization. 

Some insight into the polarization switching behaviour of spin-VCSELs can be gained by considering first the steady-state solutions (equilibria) of the SFM equations for elliptically-polarized fields. These are characterised by a constant phase difference between the LCP and RCP components of the optical field \cite{r7,r12}. For the case of linearly-polarised (LP) pumping, when this phase difference is $0$ the VCSEL output is LP with the field in the $x$-direction (the in-phase mode); a phase difference of $\pi$ gives LP emission with the field in the $y$-direction (the out-of-phase mode). For elliptically polarised pumping the lasing emission is, in general, elliptically polarised with two solutions corresponding to the cases when the phase difference is the “continuation” either of $0$ or $\pi$; hence we refer to these two cases as “in-phase” or “out-of-phase” solutions. The aim of this work is to explain why the spin-VCSEL system chooses one solution over the other for a given operating condition.

The only stability analysis to have been reported (to our knowledge) is for the case of LP pumping where the SFM equations can be studied by perturbing around the LP modes \cite{r16,r5,1a,1aa,3a,2a,4a,5a,6a,7a,8a,9a}. The stability analysis of the LP solutions provides a system of equations that decouple (in the linear approximation) into two subsets, each of three coupled equations. The first subset describes the fluctuations of the LP fields and the total electron density; a pair of eigenvalues determines the frequency and damping of the relaxation oscillations which are controlled by some parameters and are a well-known feature of laser dynamics. This demonstrates that the LP modes are stable with respect to perturbations by amplitude perturbations of the same polarisation. The remaining eigenvalue is zero and is associated with the arbitrariness of the phase of the electric field. The second subset of equations characterises the stability of a polarised solution with respect to perturbations of the orthogonal polarisation. This yields a third order characteristic polynomial, analysis of which produces various regimes of dynamics including polarisation oscillations. Polarisation switching between the LP modes has also been discussed for this case \cite{r5}; algebraic results for borders separating regions of LP mode stability have been obtained \cite{r5,3a,5a,4a}. 
However, no systematic stability analysis has been reported for the case of non-vanishing optical pump ellipticity, which we provide here. 

After an initial discussion of the SFM equations, we present first a small-signal (asymptotic) stability analysis for the case of LP optical pumping just above lasing threshold. Analytical results are obtained for the stability of both the in-phase and out-of-phase solutions. Next the small-signal analysis is extended to the case of very small optical pump elipticity, and again asymptotic analytical results are obtained for both solutions. These analytical results are then compared with numerical computations of the eigenvalues of the SFM system, revealing good agreement for a typical set of values of the spin-VCSEL parameters. In addition numerical results are presented for the output polarization versus the pumping polarization for much higher values of optical pumping above threshold and for the full range of pumping polarization ellipticity (from linear up to circular). Finally some experimental results of the ellipticity behaviour are presented and interpreted in the context of the theory in terms of changes of stability between in-phase and out-of-phase solutions.

\section{Spin-laser model and time-independent solutions}

In the SFM \cite{r16}, the circularly polarised electric field components are coupled by the crystal birefringence, characterised by a rate $\gp$. Gain anisotropy (dichroism) due to cavity geometry and other effects is also included with a rate $\ga$. Thus the complex rate equations for the time-dependence of the right- and left-circularly polarised field components, denoted by $\Ep$ and $\Em$, respectively, are
\begin{align}
\frac{d\Ep}{dt}&=\kappa\(N+m-1\)\(1+i\alpha\)\Ep-\(\ga+i\gp\)\Em,\label{gov1a}\\
\frac{d\Em}{dt}&=\kappa\(N-m-1\)\(1+i\alpha\)\Em-\(\ga+i\gp\)\Ep\label{gov1b}
\end{align}
where $\k$ is the cavity decay rate and $\a$ is the 'linewidth enhancement factor' that relates changes in real and imaginary part of the refractive index.

The normalised carrier variables $N$ and $m$ appearing in \eqref{gov1a} and \eqref{gov1b} are defined by $N = (n_+ + n_-)/2$ and $m = (n_+ - n_-)/2$, where $n_+$ and $n_-$ are the corresponding normalised densities of electrons with spin-down and spin-up, respectively. The rate equations for these variables are \cite{r7}
\begin{align}
\frac{dN}{dt}&=\g\ls\eta-\(1+|\Ep|^2+|\Em|^2\)N\right.\nonumber\\
&\left.-\(|\Ep|^2-|\Em|^2\)m\rs,\label{gov1c}\\
\frac{dm}{dt}&=\g P\eta -\ls\gs+\g\(|\Ep|^2+|\Em|^2\)\rs m\nonumber\\
&-\g\(|\Ep|^2-|\Em|^2\)N,\label{gov1d}
\end{align}
where $\g$ is the electron density decay rate,  $\gs$ is the spin relaxation rate, $\eta=\eta_++\eta_-$ is the total normalised pump power and the pump polarisation ellipticity $P$ is defined as
\begin{equation}
P=\frac{\eta_+-\eta_-}{\eta_++\eta_-},
\label{P}
\end{equation}
where $(\eta_+,\eta_-)$ are dimensionless circularly-polarised pump components that describe polarised optical pumping.

The SFM equations \eqref{gov1a}-\eqref{gov1d} are quite general in the spin-polarised pumping terms and can equally well apply to electrical pumping as to optical pumping \cite{r1}.

The spin-laser output is usually expressed in terms of circularly polarised intensities $I_+ = |\Ep|^2$, $I_- = |\Em|^2$, $I_{total} = (I_+ + I_-)$, and polarisation ellipticity $\epsilon$ defined as
\begin{equation}
\epsilon=\frac{I_+-I_-}{I_++I_-}.
\label{eps}
\end{equation}
Values of $P$ or $\epsilon$  of $+1 (-1)$ correspond to right (left) circular polarisation, whilst a value of 0 corresponds to linear polarisation. Note that the equation is invariant under the transformation $P\to-P$, $m\to-m$, $E_\pm\to E_\mp$. Therefore, without loss of generality one may only consider the case of $P>0$.

Our analysis is particularly pertinent to time-independent solutions. In that case, we look for solutions in a rotating frame of the form
\begin{equation}
\Ep=\ep\,e^{i\omega t},\,\Em=\em\,e^{i\theta}\,e^{i\omega t},\,N=N_s,\,m=m_s,
\label{sts}
\end{equation}
with all the unknown variables, i.e., $\ep,\,\em,\,\theta,\,\w,\,N_s,\,m_s$ being time-independent and real-valued. When $\theta$ is the "continuation" of 0 or $\pi$, we refer to the solution as in-phase or out-of-phase, respectively.

The linear stability of the time-independent solution is obtained by substituting $\Ep=\(\ep+\varepsilon\widehat{E}_+e^{\lambda t}\)e^{i\w t},\,\Em=\(\em e^{i\theta}+\varepsilon\widehat{E}_-e^{\lambda t}\)e^{i\omega t},\,N=N_s+\varepsilon\widehat{N}e^{\lambda t},\,m=m_s+\varepsilon\widehat{m}e^{\lambda t}$ into the governing equations and linearising for small $\varepsilon$ to obtain the eigenvalue problem
\begin{equation}
\M\v=\l\v,
\label{evp}
\end{equation}
where $\v=\(\widehat{E}_+,\widehat{E}_-,\widehat{E}_+^*,\widehat{E}_-^*,\widehat{N},\widehat{m}\)^T$, $\mathbf{a}^T$ denotes the transpose of the matrix $\mathbf{a}$, $\star$ represents complex conjugation, and
\begin{widetext}
\begin{equation}
\mathcal{M} =
\(\begin{array}{cccccc}

M_{11} & M_{12} & 0 & 0 & K_1\ep & K_1\ep\\
M_{12} & M_{22} & 0 & 0 & K_1\em e^{i\theta} & -K_1\em e^{i\theta}\\
0 & 0 & M_{11}^* & M_{12}^* & K_1^*\ep & K_1^*\ep\\
0 & 0 & M_{12}^* & M_{22}^* & K_1^*\em e^{-i\theta} & -K_1^*\em e^{-i\theta}\\
K_2\ep & K_3\em e^{-i\theta} & K_2\ep & K_3\em e^{i\theta} & M_{55} & M_{56}\\
K_2\ep & -K_3\em e^{-i\theta} & K_2\ep & -K_3\em e^{i\theta} & M_{56} & M_{66}
\end{array}\)
\label{M}
\end{equation}
with
\begin{align*}
M_{11}&=\k(N_s+m_s-1)(1+i\a)-i\w,\,M_{12}=-(\ga+i\gp),\\
M_{22}&=\k(N_s-m_s-1)(1+i\a)-i\w,\\
M_{55}&=-\g\(1+\ep^2+\em^2\),\,M_{56}=-\g\(\ep^2-\em^2\),\,M_{66}=-\(\gs+\g\(\ep^2+\em^2\)\),\\
K_1&=\k(1+i\a),\, K_2=-\g(N_s+m_s),\,K_3=\g(-N_s+m_s).
\end{align*}
\end{widetext}
It is clear that the solution is unstable when there is an eigenvalue with Re$(\lambda)>0$ and stable when Re$(\lambda)<0$.

\section{Vanishing pump polarisation ellipticity: P=0}
\label{secp0}

First, consider the case of linear polarisation $P=0$. One can check that \cite{r5,4a}
\begin{align}
\dis\ep&=\em=\sqrt{\frac{\eta_1}{2N_s}},\, N_s=1+\frac{\ga}\k\cos\theta,\\
\dis&\w\cos\theta=\ga\a-\gp,\, m_s=0,
\label{sol1}
\end{align}
where $\eta_1=\eta-N_s$ and $\theta=0,\pi$ are time-independent solutions of the governing equations.

The stability of LP modes in the general case $\eta_1=\mathcal{O}(1)$ has been considered in \cite{4a}. However, no explicit expression of the eigenvalues is presented, which will be needed later for the case of $P\neq0$. Here, we will study the stability analytically for $0<\eta_1\ll1$ and assume that the other parameters are $\mathcal{O}(1)$. It is therefore natural to expand the variables in the eigenvalue problem \eqref{evp} as the followings
\begin{align}
\dis\M&=\M_{0,0}+\sqrt{\eta_1}\M_{0,1}+{\eta_1}\M_{0,2}+\dots,\nonumber\\
\dis\v&=\v_{0,0}+\sqrt{\eta_1}\v_{0,1}+{\eta_1}\v_{0,2}+\dots,\\
\dis\l&=\l_{0,0}+\sqrt{\eta_1}\l_{0,1}+{\eta_1}\l_{0,2}+\dots.\nonumber
\end{align}

\subsection{Stability of in-phase solutions}

When $\theta=0$, we obtain that
\begin{widetext}
\begin{align}
\dis\M_{0,0}&=
\(\begin{array}{cccccc}
\ga+i\gp & -\ga-i\gp & 0 & 0 & 0 & 0\\
-\ga-i\gp & \ga+i\gp & 0 & 0 & 0 & 0\\
0 & 0 & \ga-i\gp & -\ga+i\gp & 0 &  0\\
0 & 0 & -\ga+i\gp & \ga-i\gp & 0 &  0\\
0 & 0 & 0 & 0& -\g & 0\\
0 & 0 & 0 & 0 & 0& -\gs
\end{array}\),\\
\dis\M_{0,1}&=
\(\begin{array}{cccccc}
0 & 0 & 0 & 0& \frac{\k(1+i\a)}{\sqrt{2N_s}} & \frac{\k(1+i\a)}{\sqrt{2N_s}}\\
0 & 0 & 0 & 0& \frac{\k(1+i\a)}{\sqrt{2N_s}} & \frac{-\k(1+i\a)}{\sqrt{2N_s}}\\
0 & 0 & 0 & 0& \frac{\k(1-i\a)}{\sqrt{2N_s}} & \frac{\k(1-i\a)}{\sqrt{2N_s}}\\
0 & 0 & 0 & 0& \frac{\k(1-i\a)}{\sqrt{2N_s}} & 0\\
-\g\sqrt{\frac{N_s}{2}} & -\g\sqrt{\frac{N_s}{2}} & -\g\sqrt{\frac{N_s}{2}} & -\g\sqrt{\frac{N_s}{2}} & 0 & 0\\
-\g\sqrt{\frac{N_s}{2}} & \g\sqrt{\frac{N_s}{2}} & -\g\sqrt{\frac{N_s}{2}} & \g\sqrt{\frac{N_s}{2}} & 0 & 0
\end{array}\),\\
\dis\M_{0,2}&=
\(\begin{array}{cccccc}
0 & 0 & 0 & 0& 0 & 0\\
0 & 0 & 0 & 0& 0 & 0\\
0 & 0 & 0 & 0& 0 & 0\\
0 & 0 & 0 & 0& 0 & 0\\
0 & 0 & 0 & 0& -\g/N_s & 0\\
0 & 0 & 0 & 0& 0 & -\g/N_s\\
\end{array}\).
\end{align}
\end{widetext}

From \eqref{evp}, terms at $\mathcal{O}(1)$ yield
\be
\(\M_{0,0}-\l_{0,0}\)\v_{0,0}=0,
\label{e1}
\ee
from which we obtain the eigenvalues
\be
\l_{0,0}=0,\,2\(\ga\pm i\gp\),\,-\gs,\,-\g.
\ee
The eigenvalue $\l_{0,0}=0$ has double algebraic and geometric multiplicity, with one of them is due to the gauge phase invariance of the governing equations \eqref{gov1a}--\eqref{gov1d}.

When $\eta_1$ is switched on, the only source of instability is any eigenvalue with vanishing real part. It is therefore necessary to track the influence of the parameter on the eigenvalue. In addition to the zero eigenvalues, we will also need to compute the bifurcation of the eigenvalues $\l_{0,0}=2\left(\ga\pm i\gp\right)$ particularly because for our experimental set-up, the gain anisotropy $\ga$ is negligibly small.

\subsubsection{$\l_{0,0}=0$}

The corresponding eigenvectors of the eigenvalue are
\be
v_1=\(\begin{array}{c}
1\\1\\0\\0\\0\\0
\end{array}\),\,
v_2=\(\begin{array}{c}
0\\0\\1\\1\\0\\0
\end{array}\).
\label{v12}
\ee
One therefore obtains that a generalised corresponding eigenvector of the eigenvalue is
\be
\v_{0,0}=c_1v_1+c_2v_2,
\ee
with $c_j$ being a constant.

Terms at $\mathcal{O}(\sqrt{\eta_1})$ give us
\be
\(\M_{0,0}-\l_{0,0}\)\v_{0,1}=\(\l_{0,1}-\M_{0,1}\)\v_{0,0}.
\label{e2}
\ee
As the matrix operator $\(\M_{0,0}-\l_{0,0}\)$ on the l.h.s.\ of the equation is the same as \eqref{e1}, \eqref{e2} can have a solution provided that the r.h.s.\ is orthogonal to the null-space of the Hermitian (conjugate) transpose of the matrix operator, i.e.\ $\(\M_{0,0}-\l_{0,0}\)^H$. The orthogonality is with respect to the common inner product
\[
<\mathbf{a},\mathbf{b}>=\mathbf{b}^H\mathbf{a}.
\]

Here, one can easily compute that the null-space of $\(\M_{0,0}-\l_{0,0}\)^H$ are spanned by $v_1$ and $v_2$ \eqref{v12} from which we obtain that $\l_{0,1}=0$ and
\be
\v_{0,1}=\(\begin{array}{c}
0\\0\\0\\0\\-\sqrt{2N_s}\(c_1+c_2\)\\0
\end{array}\),
\label{v2}
\ee
from solving \eqref{e2}.

At the order $\mathcal{O}({\eta_1})$, we have the system
\be
\(\M_{0,0}-\l_{0,0}\)\v_{0,2}=\(\l_{0,2}-\M_{0,2}\)\v_{0,0}-\M_{0,1}\v_{0,1}.
\ee
Applying the same procedure as before, we obtain the coupled equations
\be
\lambda_{0,2} c_1 = (i\a-1) \k (c_1 + c_2),\,
\lambda_{0,2}c_2 = -( i\a+1) \k (c_1 + c_2).
\ee
Solving the coupled equations as an eigenvalue problem yields
\be
\l_{0,2}=0,\,-2\k.
\ee
Therefore, we obtain that one of the zero eigenvalues bifurcates linearly for small $\eta_1$ as
\be
\l=-2\k\eta_1+\mathcal{O}\(\eta_1^{3/2}\).
\label{ev00}
\ee

\subsubsection{$\l_{0,0}=2\(\ga\pm i\gp\)$}

Here, we only consider one of the eigenvalue pair, i.e.\ $\l_{0,0}=2\(\ga+ i\gp\)$. The corresponding eigenvector of the eigenvalue is
\be
\v_{0,0}=\(
\begin{array}{c}
-1\\1\\0\\0\\0\\0
\end{array}
\).
\ee

\be
\v_{0,1}=\(
\begin{array}{c}
0\\0\\0\\0\\0\\\displaystyle\frac{\g\sqrt2\sqrt{(\k+\ga)\k}}{\k(2i\gp+\gs+2\ga)}
\end{array}
\).
\ee

Following the same procedure as above, we obtain that 
\be
\l_{0,2} = -\frac{(1+i\a)\k\g}{2\ga+2i\gp+\gs}.
 \ee
Thus, the eigenvalue bifurcates linearly as $\eta_1$ is increased.

\subsubsection{Other eigenvalues}

For the sake of completeness, using the same analysis we obtain that the other eigenvalues bifurcate as
\begin{align}
\lambda&=-\g+2\,\eta_1\(\ga+\k-\g/2\)/N_s+\dots,\\
\lambda&=-\gs+\g\k\eta_1\times\left[4\gp\(\ga\alpha+\gp+\k\alpha\)+\right.\nonumber\\
&\left.2\ga\(4\ga+2\k+3\gs\)+\gs\(2\k+\gs\)\right]/\nonumber\\
&\left[\(4\gp^2+4\ga^2+4\gs\ga+\gs^2\)\(\k+\ga\)\right]+\dots.
\end{align}
Note that these eigenvalues are initially on the left half plane and hence cannot create instability for small $\eta_1$.

\subsection{Stability of out-of-phase solutions}

One can do the same calculations as above. Therefore, here we will only present our results. The eigenvalues of the time-independent solution \eqref{sol1} for $\theta=\pi$ and small $\eta_1$ are given by
\begin{align}
\lambda&=0,\,-2\k\eta_1+\dots,\,-\g-2\eta_1\(\ga -\k + \g/2\)/N_s+\dots,\nonumber\\
&-2\(\ga\pm i\gp\)-\eta_1(1\pm i\alpha)\k\g/(2\ga\pm2i\gp-\gs)+\dots,\nonumber\\
&-\gs-\eta_1\k\g\times\left[4\gp\(\ga\alpha+\gp-\k\alpha\)+\right.\nonumber\\
&\left.2\ga\(4\ga-2\k-3\gs\)+\gs\(\gs+2\k\)\right]/\nonumber\\
&\left[(4\gp^2+4\ga^2-4\gs\ga+\gs^2)\(-\k+\ga\))\right]+\dots.
\label{evpi}
\end{align}

\begin{figure*}[htbp!]
\center\subfigure{\hspace{-1cm}\includegraphics[scale=0.6,clip=]{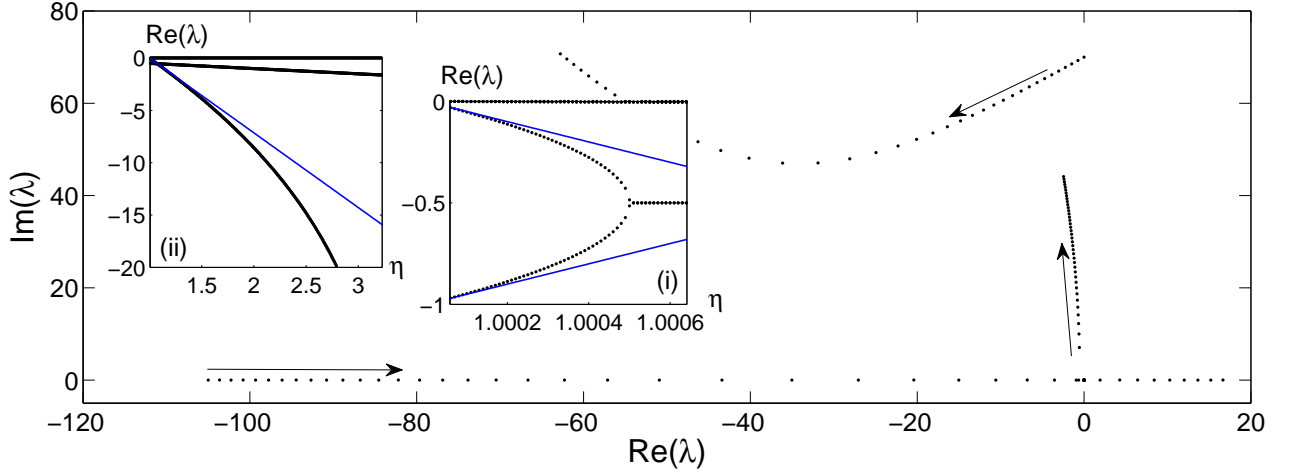}}
\caption{(Color online) The eigenvalues of the in-phase solution in the complex plane as $\eta$ increases from 1, with the trajectory direction indicated by the arrows. The insets compare some of the numerically obtained eigenvalues (dots) and our analytical approximations (solid blue curves).}
\label{fig1}
\end{figure*}

\section{Nonvanishing pump polarisation}

Next, we consider the existence and stability of the time-independent solutions when $P\neq0$. In particular, we study analytically the case of $0<\eta_1,P\ll1$ and assume that the other parameters are $\mathcal{O}(1)$. One would expect that the computation will be similar as before. However, it is important to note that here we have two small parameters which can be competing. In the following, our analysis is formal and we assume that the series is convergent. 

\subsection{In-phase solutions}

The asymptotic expansions of the in-phase solutions can be written as
\begin{align}
\ep&=\sqrt{\frac{\eta_1}{2\(1+\frac{\ga}\k\)}}+{\ep}_1P+{\ep}_2P^2+\dots,\nonumber\\
\em&=\(\sqrt{\frac{\eta_1}{2\(1+\frac{\ga}\k\)}}+{\em}_1P+{\em}_2P^2+\dots\)e^{i\(\theta_1P+\dots\)},\nonumber\\
N_s&={1+\ga/\k}+N_2P^2+\dots,\,\w=\ga\a-\gp+\w_2P^2+\dots,\nonumber\\ m_s&=m_1P+\dots.
\label{sol2}
\end{align}

Performing perturbation expansions as before but now in $P$, we obtain
\begin{align}
{\ep}_1&=-{\em}_1= -\frac14\frac{\sqrt{2\eta_1}\k\a\g}{\gp\gs}+\mathcal{O}\(\eta_1^{3/2},\ga\sqrt\eta_1\),\\
{\ep}_2&={\em}_2=\frac{-\a\gp\k\g^2}{\sqrt{8\eta_1}\gp^2\gs^2}+\mathcal{O}\(\sqrt{\eta_1},\ga/\sqrt{\eta_1}\),\\
\theta_1&=\frac{-\g\k}{\gp\gs}+\mathcal{O}\(\ga\),\,\w_2=\frac12\frac{(\a^2+1)\g^2\k^2}{\gs^2
\gp}+\mathcal{O}\(\ga\),\\ 
m_1&=\g(\k+\ga)/(\k\gs),\,N_2=\frac{\g^2\k\a}{\gs^2\gp}+\mathcal{O}\(\ga\).
\end{align}
Note that ${\ep}_2={\em}_2$ becomes singular in the limit $\eta_1\to0$. This informs us that the expansion \eqref{sol2} is valid provided that $P^2\ll\sqrt{\eta_1}$ and there may be bifurcations when this condition is violated.

Next, we study the stability of the solutions. It is natural to expand the variables in the eigenvalue problem \eqref{evp} as
\be
\square=\square_0+\square_1P+\square_2P^2+\dots,
\ee
where $\square=\M,\,\v,\,\l$. Substituting the expansion in the eigenvalue problem, we obtain at $\mathcal{O}(1)$, $\mathcal{O}(P)$, and $\mathcal{O}(P^2)$, respectively
\begin{align}
&\(\M_0-\l_0\)\v_0=0,\,\(\M_0-\l_0\)\v_1=\(\l_1-\M_1\)\v_0,\nonumber\\
&\(\M_0-\l_0\)\v_2=\(\l_2-\M_2\)\v_0+\(\l_1-\M_1\)\v_1.
\end{align}
Note that the equation at $\mathcal{O}(1)$ is the same as that solved in the previous section. Therefore, we will expand each variable in $\eta_1$ and will solve the corresponding eigenvalue problems asymptotically, i.e., we write for $\square_j$, $j=0,1,2,$
\begin{align}
\square_j&=\square_{j,0}+\square_{j,1}\sqrt{\eta_1}+\dots,\,j=0,1,\\
\square_2&=\frac{\square_{2,-1}}{\sqrt{\eta_1}}+\square_{2,0}+\dots.
\end{align}
Due to the expansion, it can be easily checked that the asymptotic values of $\square_0$ will be the same as those obtained in Section \ref{secp0} above. 

First, consider the eigenvalue 
\[
\l_{0,0}=2\(\ga \pm i\gp\).
\]
From the equation at order $\mathcal{O}(P,\eta_1^0)$, i.e., 
\[
\(\M_{0,0}-\l_{0,0}\)\v_{1,0}=\(\l_{1,0}-\M_{1,0}\)\v_{0,0},
\]
its solvability condition yields $\l_{1,0}=0$.

Solving the equation at order $\mathcal{O}(P^2,\eta_1^0)$, i.e., 
\[
\(\M_{0,0}-\l_{0,0}\)\v_{2,0}=\(\l_{2,-1}-\M_{2,0}\)\v_{0,0},
\]
gives us $\l_{2,-1}=0$.

A leading order non-vanishing eigenvalue in the presence of $P$ can be obtained from the equation at order $\mathcal{O}(P^2,\eta_1^1)$, i.e., 
\begin{align}
\(\M_{0,0}-\l_{0,0}\)\v_{2,1}&=\(\l_{2,0}- \M_{2,1}\)\v_{0,0}-\M_{0,1}\v_{2,0}\nonumber\\
&-\M_{1,0}\v_{1,0}-\M_{2,0}\v_{0,1},\nonumber
\end{align}
from which we obtain that up to $\mathcal{O}(\ga)$
\begin{equation}
\l_{2,0}=\frac{-(2(i\alpha-1)\gp-\alpha(\gs+\g)-i\gs)(i+\alpha)\g^2\k^2}{(i\gs+2\gp)\gp\gs^2}.
\end{equation}

For 
$$
\l_{0,0}=-\g,
$$
we obtain 
\begin{align}
\l_{2,0}=\frac{\g^2\(\g-2\k\)\a\k}{\gs^2\gp}+\mathcal{O}(\ga).
\end{align}

Performing the same calculation for 
$$
\l_{0,0}=0
$$
yields $\l_{0,1}=\l_{1,0}=\l_{2,-1}=0$ and
\begin{align}
\l_{2,0}=\frac{2\a\g^2\k^2}{\gp\gs^2}
+\mathcal{O}(\ga).
\end{align}

For 
\[
\l_{0,0}=-\gs,
\]
we obtain
\[
\lambda_{2,0}=\frac{\g^3\alpha\k\(-4\gp^2+4\alpha\gp\k-\gs^2+2\gs\k\)}{\gp\gs^2\(-4\gp^2-\gs^2\)}
\]

\subsection{Out-of-phase solutions}

The asymptotic expressions of the out-of-phase solutions are written as
\begin{align}
\ep&=\frac{\sqrt\eta_1}{\sqrt{2\(1-\ga/\k\)}}+{\ep}_1P+{\ep}_2P^2+\dots,\nonumber\\
\em&=\(\frac{-\sqrt\eta_1}{\sqrt{2\(1-\ga/\k\)}}+{\em}_1P+{\em}_2P^2+\dots\)e^{i\(\theta_1P+\dots\)},\nonumber\\
N_s&=\(1-\ga/\k\)+\dots,\,\w=\gp-\ga\a+\dots,\nonumber\\ m_s&=m_1P+\dots.
\label{solo2}
\end{align}

Performing perturbation expansions as before, we obtain
\begin{align}
{\ep}_1&={\em}_1= \frac14\frac{\sqrt{2\eta_1}\k\g\alp}{\gp\gs}+\mathcal{O}\(\eta_1^{3/2},\ga\sqrt\eta_1\),\\
{\ep}_2&=-{\em}_2=\frac1{\sqrt{8}}\frac{\alp\g^2\k}{\sqrt\eta_1\gp\gs^2}+\mathcal{O}\(\sqrt{\eta_1},\ga/\sqrt\eta_1\),\\
N_2&=-\frac12\frac{2\k\alp\g^2}{\gp\gs^2}+\mathcal{O}(\ga),\\
w_2&=-\frac12\frac{\k^2(\alp^2+1)\g^2}{\gp\gs^2}+\mathcal{O}(\ga),\\
\theta_1&=\frac{\k\g}{\gp\gs}+\mathcal{O}(\ga),\,
 m_1=\g(\k-\ga)/(\k\gs).
\end{align}
Note that ${\ep}_2={\em}_2$ also becomes singular in the limit $\eta_1\to0$. 

Next, we study the stability of the solutions. Using the same expansions and following the same procedures as above, we obtain that for the non-zero eigenvalue
\[
\l_{0,0}=-2\(\ga+i\gp\),
\]
the pump yields the correction
\[
\l_{2,0}=\frac{(2(i\a+1)\gp+i\gs-(\g+\gs)\a)(i-\a)\g^2\k^2}{(i\gs+2\gp)\gp\gs^2}.
\]

For 
$$
\l_{0,0}=-\g,
$$
we obtain 
\begin{align}
\l_{2,0}=-\frac{(\g-2\k)\alpha\k\g^2}{\gs^2\gp}.
\end{align}

For $\l_{0,0}=0$, we also obtain 
\[
\l_{2,0}=-\frac{2\alpha\k^2\g^2}{\gp\gs^2}.
\]

For $\l_{0,0}=-\gs$, we obtain
\[
\l_{2,0}=-\frac{\g^3\alpha\k\(-2\gs\k+4\alpha\gp\k+\gs^2+4\gp^2\)}{\gp\gs^2\(\gs^2+4\gp^2\)}.
\]

\section{Numerical results}

\begin{figure*}[hbpt!]
\center\subfigure{\hspace{-1cm}\includegraphics[scale=0.6,clip=]{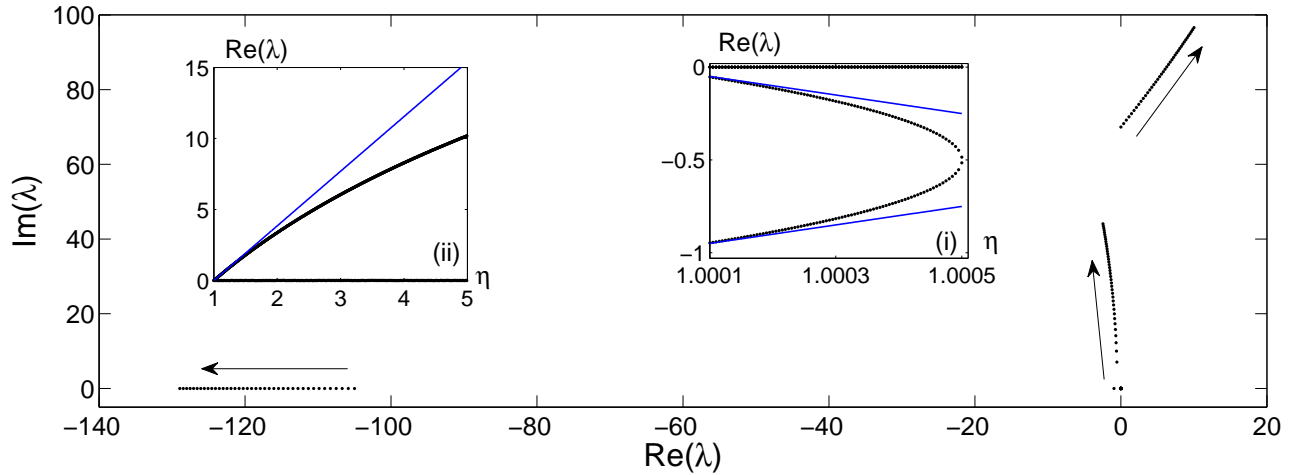}}
\caption{(Color online) The same as Fig.\ \ref{fig1}, but for the out-of-phase solution}
\label{fig2}
\end{figure*}

We solved the governing equations \eqref{gov1a}-\eqref{gov1d}, \eqref{sts} numerically using a Newton-Raphson method. To track the solution continuation when there is a saddle-node bifurcation, we used a pseudo-arclength method. The stability of the solution is then determined by solving the eigenvalue problem \eqref{evp}. 

In the following, we take the linewidth enhancement factor $\alpha = 5$, birefringence rate $\gp = 35$ ns$^{-1}$, spin relaxation rate $\gs = 105$ ns$^{-1}$, dichroism rate $\ga = 0$, carrier recombination rate $\g = 1$ ns$^{-1}$, and the cavity decay rate $\k = 250$ ns$^{-1}$.

Shown in Fig.\ \ref{fig1} are the eigenvalues $\lambda$ of the in-phase solution in the upper half of the complex-plane as $\eta$ increases from $\eta=1$. 

From the figure one can conclude that in general the effect of $\eta$ on the in-phase solution is stabilizing it. This can be seen by the fact that all the eigenvalues have negative real parts as $\eta\sim 1$ varies (except the trivial eigenvalue $\lambda=0$ that is always present due to the gauge-phase invariance). 

To compare the numerics and the analytical results calculated previously, we show in inset (i) of the figure that the eigenvalues bifurcating from $0$ and $-\gamma$ collide and create a pair of complex-valued eigenvalues. Our analytical approximations are shown in blue. It is clear that the theoretical expression can only predict the dynamics of the bifurcating eigenvalues as the parameter $\eta$ is varied prior to the collision. 

We also show the dynamics of the complex eigenvalue bifurcating from $\lambda=2\(\ga \pm i\gp\)$ 
as a function of $\eta$ in the inset (ii). Depicted is the comparison between the real part of the eigenvalues computed numerically and our analytical result. It is interesting to note that the asymptotic result agrees well with the numeric in a rather large interval of $\eta$.

If small $\eta$ stabilizes the in-phase solution, large $\eta$ has the opposite effect. The in-phase solution can also be unstable for large $\eta$. The instability is due to an eigenvalue bifurcating from the far-left eigenvalue $\lambda=-\gs$. Even though we did not present a comparison with our analytical result, the bifurcation is predicted by our asymptotic expression, i.e.\ that the eigenvalue increases for increasing $\eta$. As shown in Figure \ref{fig1}, increasing $\eta$ further makes the eigenvalue originated from $\lambda=-\gs$ cross the vertical axis. This occurs at $\eta\approx4.6$. When the eigenvalue crosses the origin, our system undergoes a pitchfork bifurcation. The bifurcating solution will be addressed later. 

If $\eta\approx1$ stabilizes in-phase solutions, the parameter has the opposite effect on the out-of-phase solutions. In Fig.\ \ref{fig2} we show the behavior of the eigenvalues as $\eta$ is varied, where one can see that all the solutions are unstable. In the insets of the figure, we also show the comparison between our asymptotic and the numerical results of critical eigenvalues that potentially lead to instability, i.e.\ eigenvalues bifurcating from $\lambda=0$ and $-\g$ in inset (i) and that from $\lambda=-2\(\ga\pm i\gp\)$ in inset (ii). Again one can note the good agreement between the results. 

\begin{figure*}[htbp!]
\subfigure[]{\includegraphics[scale=0.5,clip=]{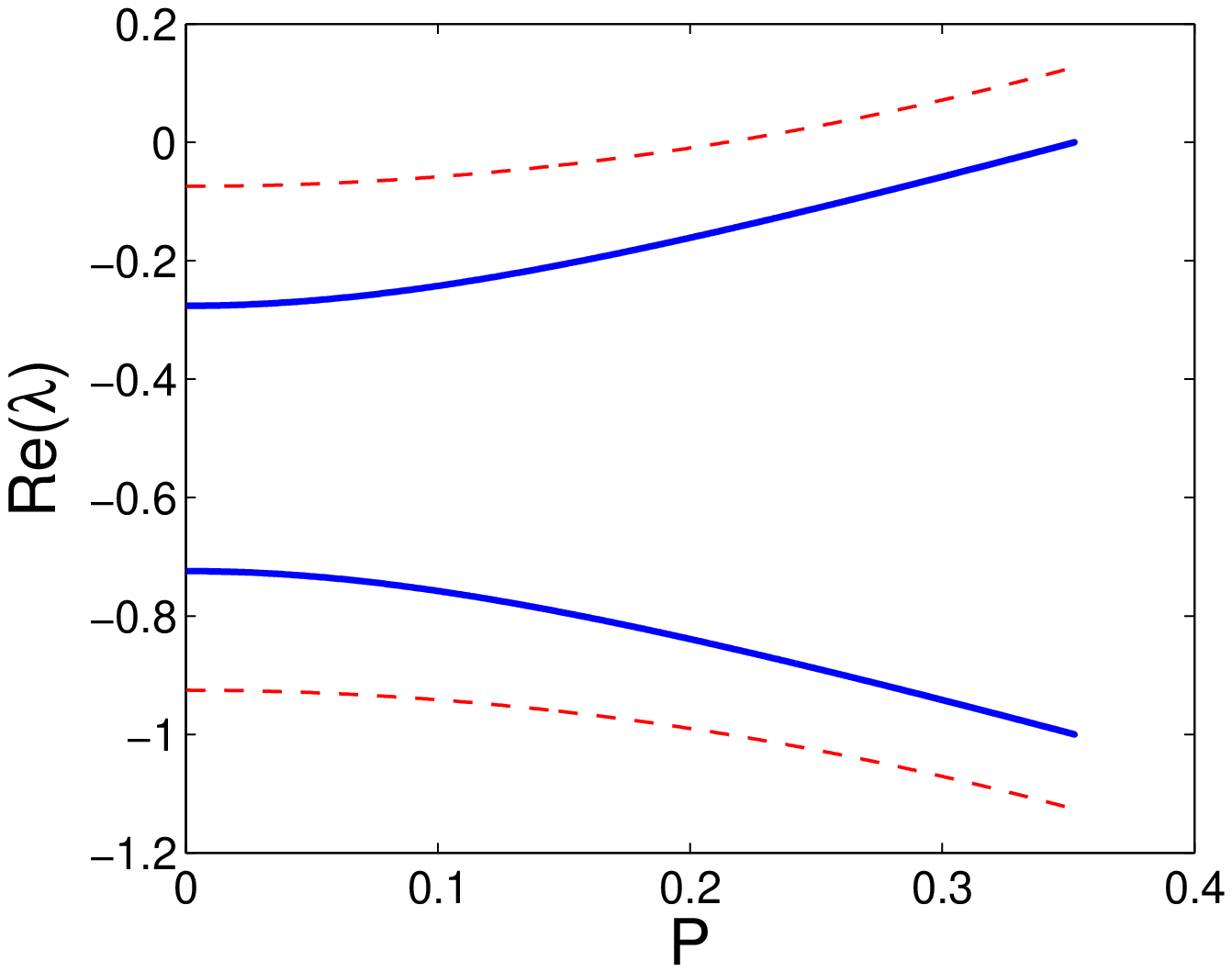}}
\subfigure[]{\includegraphics[scale=0.5,clip=]{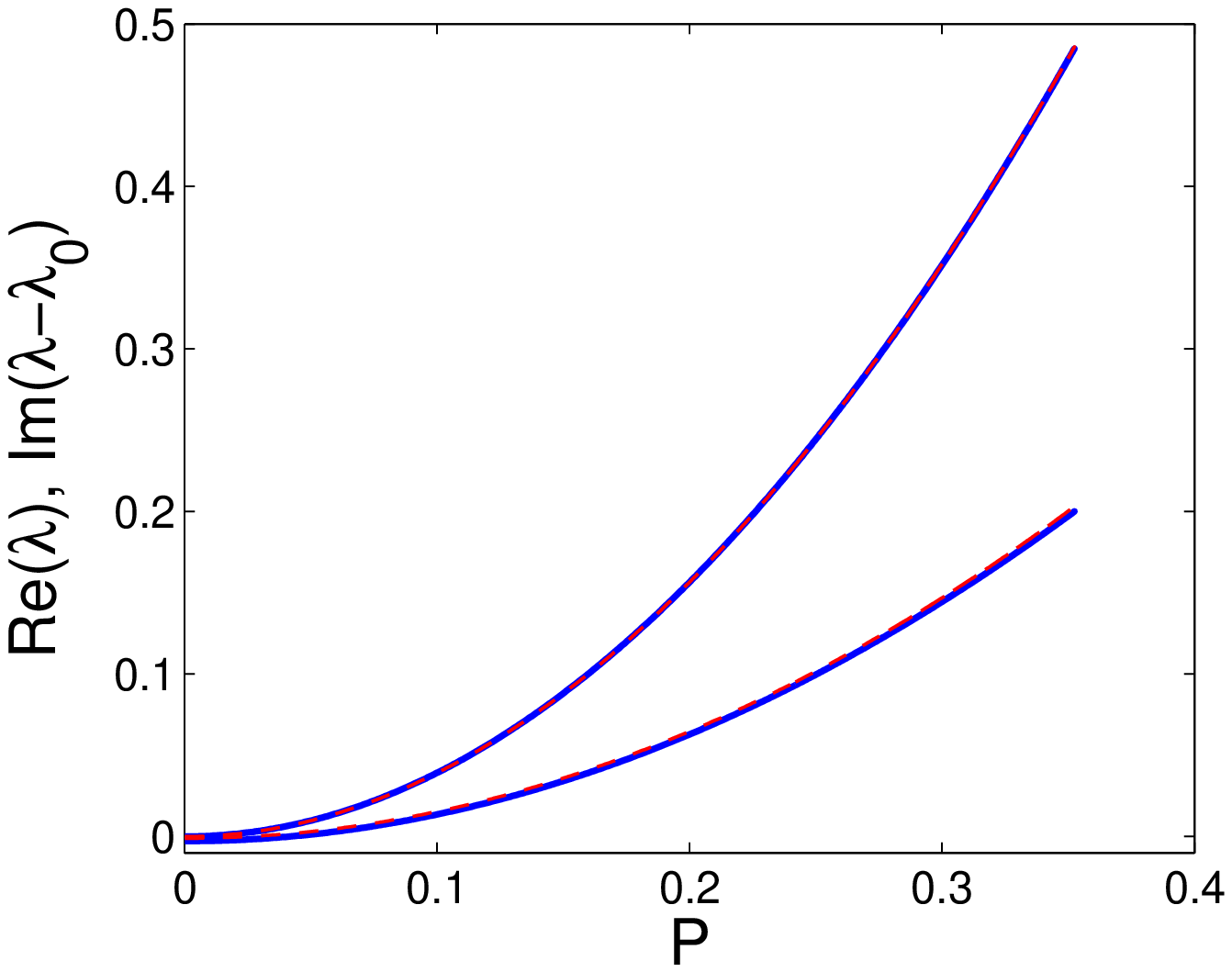}}\\
\subfigure[]{\hspace{-1cm}\includegraphics[scale=0.5,clip=]{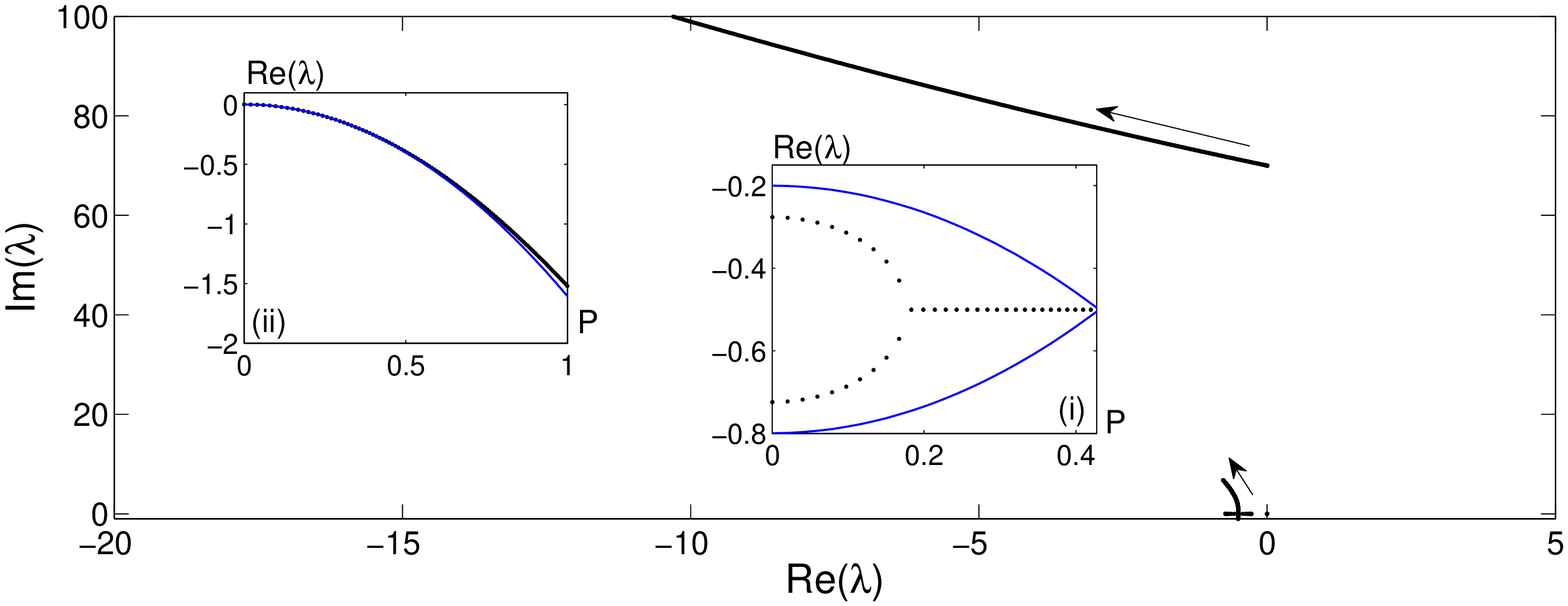}}
\center\subfigure[]{\includegraphics[scale=0.5,clip=]{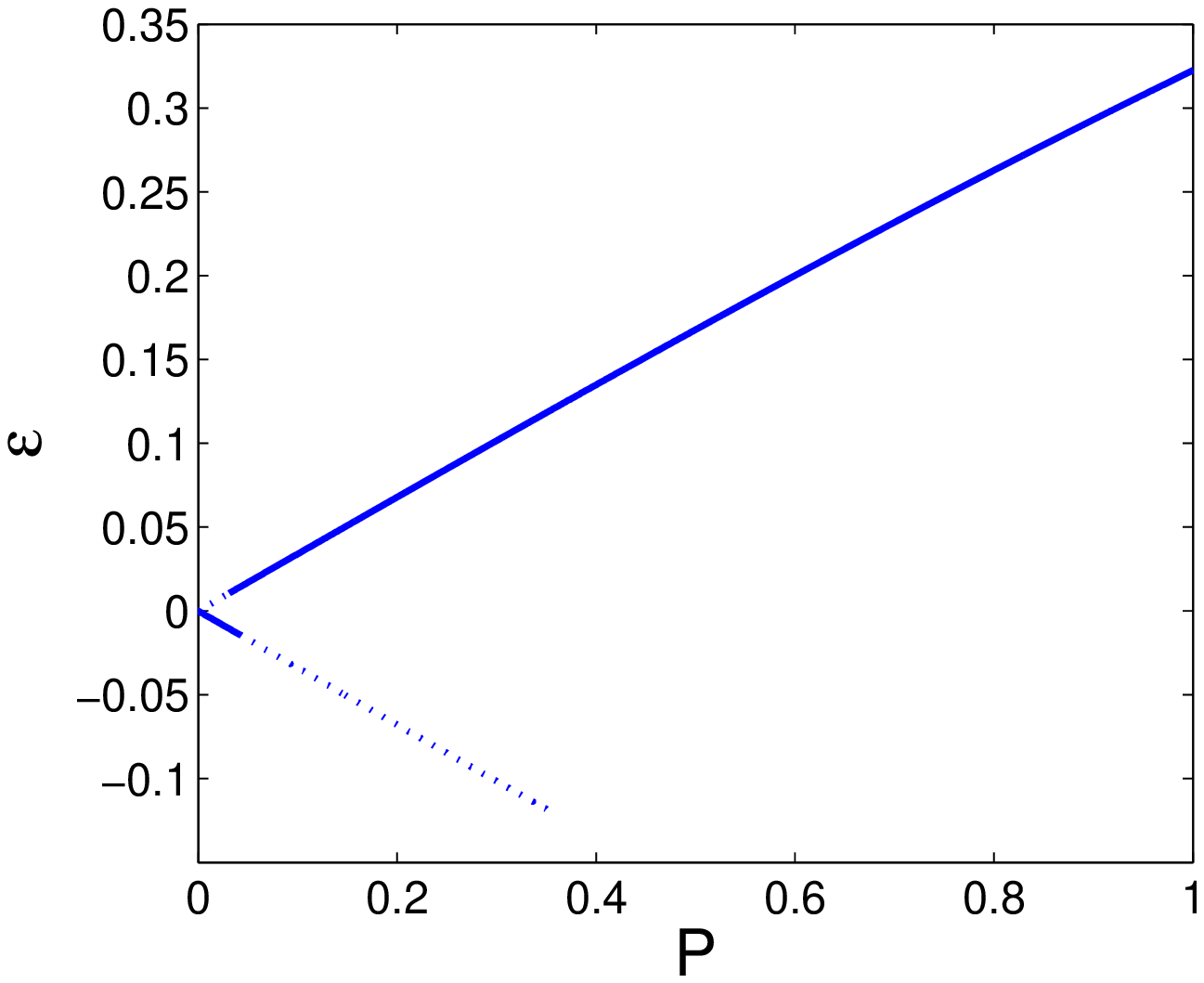}}
\caption{(Color online) (a-b) The eigenvalues of the in-phase solutions (that correspond to those in insets (i) and (ii) of Fig.\ \ref{fig1}, respectively) as a function of $P$. (c) The eigenvalues of the out-of-phase solution in the complex plane as $P$ increases from 0. (d) The ellipticity $\epsilon$ as a function of $P$. The curves with the negative and positive slope correspond to the in-phase and out-of-phase time-independent solution, respectively. Unstable and stable solutions are indicated respectively as dotted and solid lines. In all the panels, $\eta=1.0004$.}
\label{fig3}
\end{figure*}

\begin{figure*}[htbp]
\subfigure[]{\includegraphics[scale=0.6,clip=]{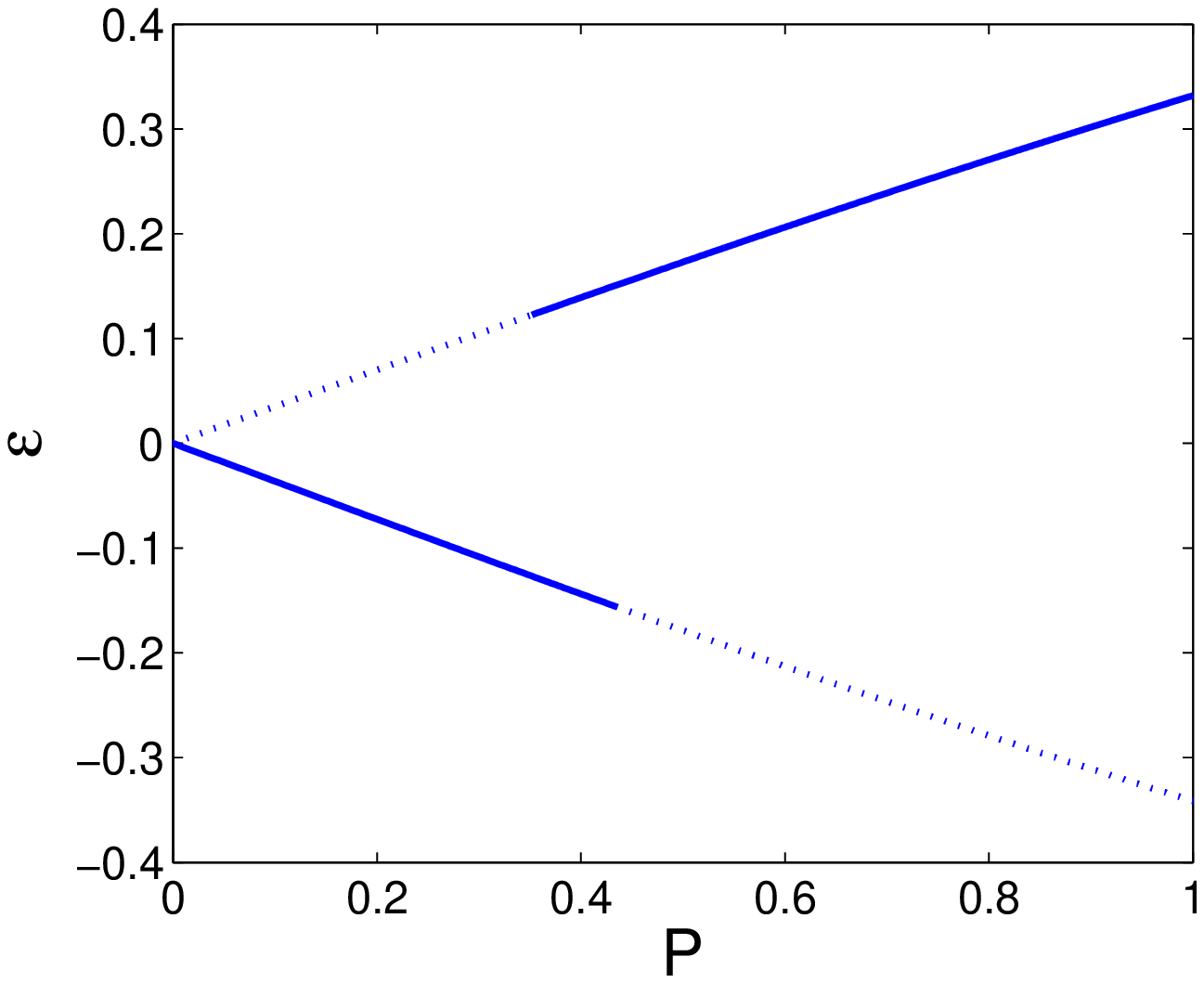}}
\subfigure[]{\includegraphics[scale=0.6,clip=]{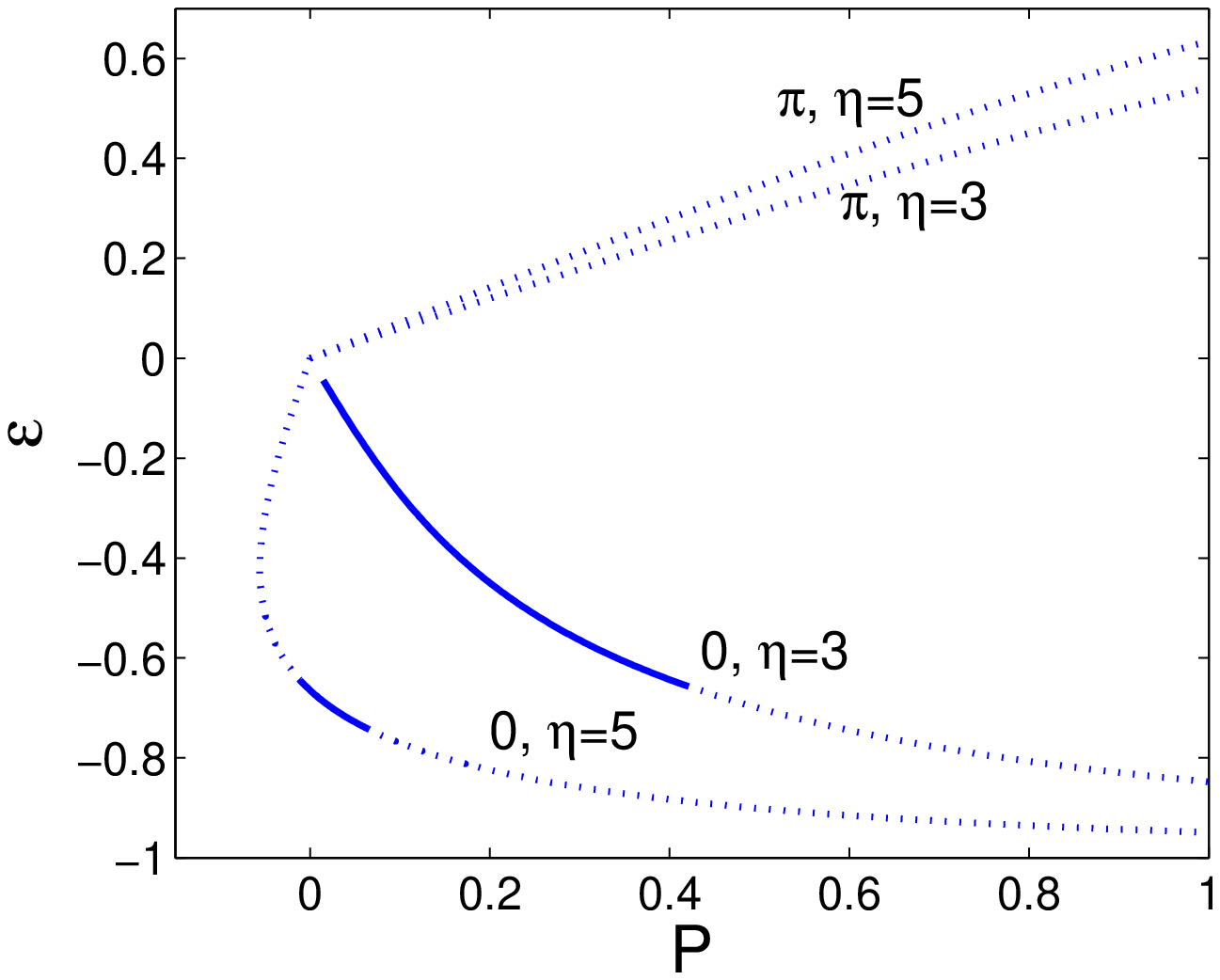}}
\caption{(Color online) The same as Fig.\ \ref{fig3}(d), but for (a) $\eta=1.05$, (b) $\eta=3$ and $5$}
\label{fig4}
\end{figure*}

Next, we consider the effect of $P$ on the stability of the in-phase and out-of-phase equilibrium solutions. 

We plot in Fig.\ \ref{fig3}(a-b) the critical eigenvalues of the in-phase solution as a function of $P$ with $\eta=1.0004$. For the two eigenvalues on the real axis that can collide and become a complex pair, our analytical result shows a qualitative agreement, where one can note that the pump polarisation tends to destabilize the solution. For the complex-valued eigenvalues that originally was on the imaginary axis, our asymptotic result shows good agreement even quantitatively as the numerical and analytical curves coincide visually. Again, it also shows that the polarisation $P\neq0$ destabilizes the solution. From combining panel (a) and (b), we found numerically that stability changes at $P=0.05$. Moreover, 
the solution ceases to exist beyond $P\approx0.35$.

In panel (c) of the same figure, we plot the eigenvalues of the out-of-phase solutions in the complex plane as $P$ varies. Our computations show that the polarisation $P\neq0$ has a stabilizing effect to the solution.  Insets (i) and (ii) in the figure present the comparison between the numerical results of the critical eigenvalues and our asymptotic analysis, where similarly to panel (a-b) we also obtain quantitative agreement for the complex pair of eigenvalues originally located at the imaginary axis. For the parameter values used in Fig.\ \ref{fig3}, we found numerically that the out-of-phase solution changes from being unstable to stable at $P=0.03$. The solution exists for any $P$. 

In Fig.\ \ref{fig3}(d), we represent the in-phase and out-of-phase solutions in terms of their ellipticity defined as \eqref{eps}. 

In Fig.\ \ref{fig3} we used the parameter value $\eta=1.0004$ for the sake of comparison with the analytical results, i.e.\ the eigenvalue bifurcating from $\lambda=0$ has not collided with another eigenvalue creating a pair of complex-valued eigenvalues. In Fig.\ \ref{fig4}, we used $\eta$ without the constraint (and hence no comparison with the analytical results). In panel (a), we still obtain the same conclusion that $P$ destabilizes the in-phase solution and stabilizes the out-of-phase one. However, the difference with Fig.\ \ref{fig3}(d) is that the in-phase and out-of-phase solutions have wider stability and instability regions, respectively. This is expected because of the effects of moderate $\eta$ to those solutions discussed previously. In addition to that, the in-phase solution also exists in a longer interval of $P$. 

However, when $\eta$ is large enough, it can destabilize the in-phase solution, see Fig.\ \ref{fig1}. We present in Fig.\ \ref{fig4}(b) examples of the case when increasing $\eta$ further does not necessarily imply a wider stability window for the in-phase solution. As the eigenvalue $\lambda$ bifurcating from $-\gs$ approaches the origin, the slope of the ellipticity curve $\epsilon(P)$ at $P=0$ is getting steeper and becomes singular at the pitchfork bifurcation. When the eigenvalue vanishes, the slope changes sign. Increasing $\eta$ further will cause the system to have another time-independent solution, i.e.\ pitchfork bifurcation, that is stable.

\section{Experimental results}

\begin{figure*}[ht]
\center\subfigure{\hspace{-1cm}\includegraphics[scale=0.5,clip=]{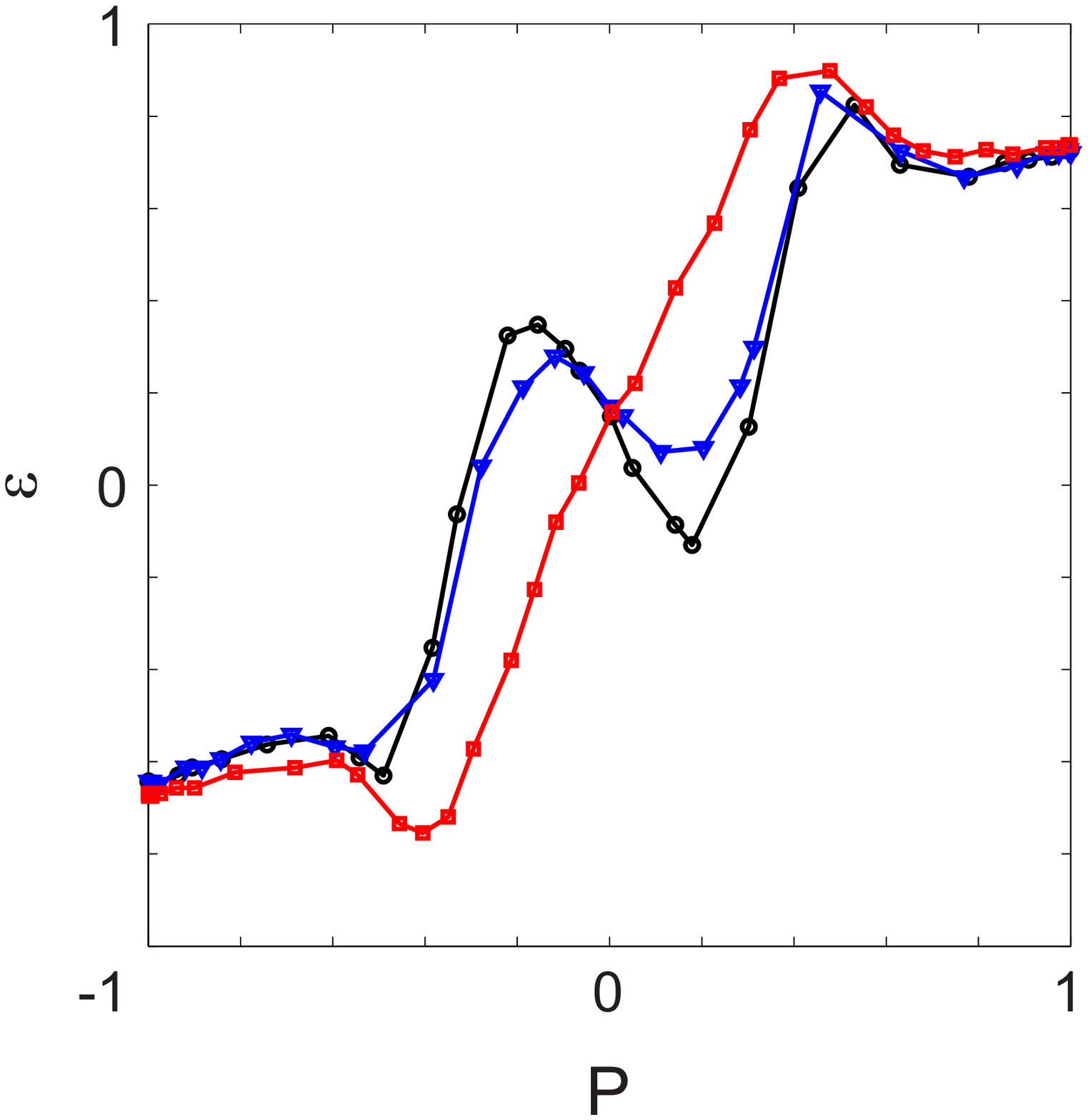}}
\caption{(Color online) Measured output polarisation ellipticity $\epsilon$ versus ellipticity of the pump laser $P$ at three different currents: 950 mA (black circles), 962 mA (blue triangles) and 1006 mA (red squares).}
\label{fig5}
\end{figure*}

The fibre-based experimental set-up has been described in detail elsewhere \cite{r4,r11,r13} and hence only a brief summary is given here. A commercial CW 980 nm laser which is controlled in terms of its polarisation and output power (via the current) is used to optically pump the VCSEL sample. The active region of the sample consists of a 3-$\lambda$ cavity that contains five groups of three GaInNAs ($\lambda$= 1300nm) quantum wells (QWs), sandwiched between high reflectivity Bragg mirror stacks; full details are given in \cite{r13}. Lasing emission from the optically pumped spin-VCSEL sample is characterised in terms of output power, wavelength, polarisation and their stability, all as a function of pump conditions.

Results for 1300 nm dilute nitride spin-VCSELs have already been reported for cases where the output showed stable lasing \cite{r13}, periodic oscillations \cite{r4} and polarisation switching \cite{r11}. Figure \ref{fig5} shows results for polarisation switching at three different pump laser currents (950mA, 962 mA and 1006 mA) above threshold (where the pump current was 875 mA). The lack of symmetry around the linearly polarised state (zero ellipticity) here arises from the fitting process used to obtain values of absolute polarisation, as discussed in \cite{r13}; in this case, differences in calibration between both polarimeters prevented optimal processing of the data and the fit was made to ensure that the extreme values of the VCSEL ellipticity are correct. Comparing these results with the theoretical ones in Figs \ref{fig3}(d) and \ref{fig4}, it is clear that there is switching between the in-phase (negative slope) and out-of-phase (positive slope) solutions (as discussed above) for each pump current. The switching always occurs from a stable branch that becomes unstable to one that is stable. The regions of stability on each branch change with pumping in the experimental results as they do for the theoretical ones. Whilst the trends are clearly similar, detailed comparison between theory and experiment is not possible at this stage since that would require more accurate knowledge of the key parameters, namely the rates of carrier recombination, spin relaxation, birefringence, dichroism and cavity decay, and the linewidth enhancement factor. Novel experimental techniques for determining these parameters in VCSELs developed recently by Perez et al \cite{r14,r15} might enable further progress in this respect.

\section{Conclusion}

We have analysed the SFM describing spin-VCSELs. In particular, we have considered the existence and stability of in-phase and out-of-phase time-independent solutions (equilibria), both in the absence and presence of pump polarisation ellipticity. For the case of LP pumping just above the lasing threshold, we showed that the in-phase solution is stable while the out-of-phase one is not. Increasing the total pump power will destabilise both types of equilibria. Additionally we showed that the pump polarisation ellipticity stabilizes the out-of-phase solution and destabilizes the other. The analytical and numerical results were shown to be in agreement qualitatively with the experiments. 

For future work, it is naturally interesting to study the attracting solutions when the system does not admit stable time-independent solutions, see Fig.\ \ref{fig4}. Normally in this region one would obtain time-periodic solutions (i.e.\ Hopf bifurcations) (see \cite{4a,9a} for the case of $P=0$). However, analytical results are currently lacking that may help understand the insight of the system for potential applications, such as information coding. 

\section*{Acknowledgement}
This work was supported by the Engineering and Physical Sciences Research Council [grant numbers EP/M024237/1 and EP/G012458/1].


\begin{thebibliography}{99.}
%





\bibitem{r1}	 N.C. Gerhardt and M.R. Hofmann, Adv. Opt. Technol. 2012, 268949 (2012)

\bibitem{ra1} M. Holub, J. Shin, S. Chakrabarti, and P. Bhattacharya, Appl. Phys. Lett. 87, 091108 (2005)

\bibitem{ra2} D. Basu, D. Saha, and P. Bhattacharya, Phys. Rev. Lett 102, 093904 (2009)

\bibitem{ra3}  J.-Y.\ Chen,	T.-M.\ Wong, C.-.\ Chang, C.-Y.\ Dong,	and Y.\-F.\ Chen, Nature Nanotech. 9, 845 (2014)

\bibitem{r2} 	R. Al-Seyab, D. Alexandropoulos, I.D. Henning and M.J. Adams, IEEE Photon. J. 3, 799 (2011)
\bibitem{r3} 	N.C. Gerhardt, M.Y. Li, H. Jahme, H. Hopfner, T. Ackemann, and M.R. Hofmann, Appl. Phys. Lett. 99, 151107 (2011)
\bibitem{r4} 	K. Schires, R. Al Seyab, A. Hurtado, V.M. Korpij\"arvi, M. Guina, I.D. Henning and M.J. Adams, IEEE Photonics Conf. (IPC), San Francisco, USA, 23-27 September 2012, pp. 870-871
\bibitem{r16} M.S. Miguel, Q. Feng and J.V. Moloney, Phys. Rev. A 52, 1728-39 (1995)
\bibitem{r5} 	J. Martin-Regalado, F. Prati, M. San Miguel and N. B. Abraham, IEEE J. Quantum Electron. 33, 765 (1997)
\bibitem{r6} 	R.K. Al-Seyab, I.D. Henning and M.J. Adams, J. Opt. Soc. Am. B 32, 683 (2015)
\bibitem{r7} 	A. Gahl, S. Balle, and M. S. Miguel, IEEE J. Quantum Electron. 35, 342 (1999) 
\bibitem{r8} 	N. Gerhardt, S. H\"ovel, M. Hofmann, J. Yang, D. Reuter and A. Wieck, Electron. Lett. 42, 88 (2006)
\bibitem{r9} 	S.S. Alharthi, R.K. Al Seyab, I.D. Henning and M.J. Adams, IET Optoelectron. 8, 117 (2014)
\bibitem{r10} 	D. Alexandropoulos, R. Al-Seyab, I.D. Henning and M.J. Adams, Opt. Lett. 37, 1700 (2012)
\bibitem{r11} 	K. Schires, R. Al Seyab, I. Henning, and M. Adams, Int. Symp. on Physics and Applications of Laser Dynamics 2013 (IS-PALD 2013), Paris, France, 29-31 October 2013
\bibitem{r12} M.J. Adams and D. Alexandropoulos, IEEE J. Quantum Electron. 45, 744 (2009)
\bibitem{r13}	 K. Schires, R. Al Seyab, A. Hurtado, V.-M. Korpij\"arvi, M. Guina, I. D. Henning and M. J. Adams, Opt. Express 20, 3550 (2012)

\bibitem{1a} M. Travagnin, M. P. van Exter, A. K. Jansen van Doorn, and J. P. Woerdman, Phys. Rev. A 54, 1647 (1996).
\bibitem{3a} J. Martin-Regalado, M. San Miguel, N.B. Abraham, and F. Prati, Opt. Lett. 21, 351 (1996).
\bibitem{2a} M. Travagnin, Phys. Rev. A 56, 4094 (1997).
\bibitem{1aa} M. Travagnin, M. P. van Exter, and J. P. Woerdman, Phys. Rev. A 56, 1497 (1997).
\bibitem{6a} J. Martin-Regalado, S. Balle, M. San Miguel, A. Valle, and L. Pesquera, Quantum Semiclassic. Opt. 9, 713 (1997).
\bibitem{5a} M.P. van Exter, A. Al-Remawi, and J.P. Woerdman, Phys. Rev. Lett. 80, 4875 (1998).
\bibitem{4a} T. Erneux, J. Danckaert, K. Panajotov, and I. Veretennicoff, Phys. Rev. A 59, 4660 (1999).
\bibitem{7a} F. Prati, P. Caccia, and F. Castelli, Phys. Rev. A 66, 063811 (2002).
\bibitem{8a} F. Prati, P. Caccia, M. Bache, and F. Castelli, Phys. Rev. A 69, 033810 (2004).
\bibitem{9a} M. Virte, K.\ Panajotov, and M.\ Sciamanna, Phys. Rev. A 87, 013834 (2013).



\bibitem{r14} P. Perez, A. Valle, I. Noriega and L. Pesquera, J. Lightwave Technol. 32, 1601 (2014)
\bibitem{r15} P. Perez, A. Valle and L. Pesquera, J. Opt. Soc. Am. B 31, 2574 (2014)



\end{thebibliography}
\end{document}